\documentstyle[12pt,epsf]{article}
\newcommand{\AAA}{{\cal A}}

\newcommand{\TT}{{\cal T}}

\newcommand{\wt}{\widetilde}
\newcommand{\wh}{\widehat}

\newcommand{\wb}{\bar}

\newcommand{\be}{\begin{equation}}
\newcommand{\ee}{\end{equation}}
\newcommand{\ben}{\begin{eqnarray}\displaystyle}
\newcommand{\een}{\end{eqnarray}}
\newcommand{\refb}[1]{(\ref{#1})}
\newcommand{\p}{\partial}
\newcommand{\sectiono}[1]{\section{#1}\setcounter{equation}{0}}

\begin{document}

{}~ \hfill\vbox{\hbox{hep-th/9902105}\hbox{MRI-PHY/P990203}
}\break

\vskip 3.5cm

\centerline{\large \bf Descent Relations Among Bosonic D-branes}
\medskip

\vspace*{6.0ex}

\centerline{\large \rm Ashoke Sen
\footnote{E-mail: asen@thwgs.cern.ch, sen@mri.ernet.in}}

\vspace*{1.5ex}

\centerline{\large \it Mehta Research Institute of Mathematics}
 \centerline{\large \it and Mathematical Physics}

\centerline{\large \it  Chhatnag Road, Jhoosi,
Allahabad 211019, INDIA}

\vspace*{4.5ex}

\centerline {\bf Abstract}

We show that the tachyonic kink solution on a pair of
D-$p$-branes in the bosonic string theory can be identified with
the D-$(p-1)$-brane of the same theory. We also speculate on the
possibility of obtaining the D-$(p-1)$-brane as a tachyonic lump
on a single D-$p$-brane. We suggest a possible reinterpretation
of the first result which unifies these two apparently different
descriptions of the D-$(p-1)$ brane.

\vfill \eject

\tableofcontents

\baselineskip=18pt

\sectiono{Introduction and Summary} \label{s1}

During the last year various relationships between the 
D-branes of
type II and type I string theories have been discovered. In
particular, it was found that
quite often a D-brane can be realised as a soliton
solution associated with the tachyon field on a brane-antibrane
pair of higher
dimension\cite{TACH,BERGABTWO,SPINOR,NOSU}. 
This fact has been used to show that
the D-branes of type II and type I string theories can be
classified by elements of the appropriate K-group of 
space-time\cite{WITTENK,HORAVAK,KONE,KTWO}.
Various other applications / generalisations of these results
have also been proposed\cite{SRED,GUTPER,PILJIN}.

In this paper we shall extend these results to the D-branes of
bosonic string theory\cite{NOTES}. 
This theory contains D-branes of all
dimensions, unlike type IIA (IIB) string theory which contains
D-branes of even (odd) dimensions only. There is a tachyonic mode
even on a single D-brane in the bosonic string theory; but when
we bring two parallel D-branes close to each other, we get extra
tachyonic modes from open strings stretched between the two
D-branes. It is this tachyonic mode which will be of interest to
us. The potential involving this tachyonic mode is even; and
according to one possible interpretation of our results, it
has a (local) minimum at non-zero value of the tachyon
field. Thus there are two degenerate minima, and
we can have tachyonic kink solution which
interpolates between the two minima. We shall show that this kink
solution describes a D-brane of one lower dimension. Thus if we
start with a pair of D-$p$-branes, the tachyonic kink solution
describes a D-$(p-1)$-brane.

Before we outline the steps involved in our analysis, let us
briefly discuss the motivation. Since bosonic string theory
itself is unstable due to the presence of the tachyon in the
closed string sector; and since the D-branes themselves are also
unstable due to the presence of the tachyon in the open string
sector, it is natural to question the usefulness of the results
of this paper.
We propose the following reasons for our study:
\begin{itemize}
\item Quite often bosonic string theory provides a simpler
setting
compared to superstring theories for the study of various issues.
Since there are several issues which are ill understood in
regarding D-branes as solitons in type II / type I string
theories, studying these issues in bosonic string theory might be
of help. One such issue involves studying the fate of the
diagonal U(1) gauge field living on the brane-antibrane pair
after tachyon condensation\cite{SRED,WITTENK}. 
Yet another issue is an explicit
construction of the soliton / vacuum configuration on the
brane-antibrane pair as a classical solution of the open string
field theory living on the pair. These issues might be simpler to
study in the bosonic string theory.

\item On a more speculative side, we would like to point out that
in view of recent developments in the subject of string duality,
it seems unlikely that the bosonic string theory will forever remain
outside the scheme that unifies all string theories as different
limits of the same underlying theory. Already some explicit
proposals have been made which relate bosonic string theory to
other theories with world-sheet supersymmetry\cite{BERGABONE}.
Study of D-branes in the bosonic string theory is certainly going
to be important if we are to study such duality relations;
although due to lack of supersymmetry our task will be much more
difficult.
\end{itemize}

The steps involved in proving the equivalence of a tachyonic kink
solution and a lower dimensional D-brane are very similar to the
ones used in
\cite{SPINOR}, where we derived a similar result for type I / type II
string theory.\footnote{Related issues have been discussed in
\cite{RECK,CAL,POL}.} However, there is a crucial difference: unlike in
\cite{SPINOR} where the boundary conformal field
theory interpolating between the $p$-brane pair and the
$(p-1)$-brane was a free fermionic field theory, in the present
case it corresponds to a level one SU(2) current
algebra theory.
We now outline the basic idea
of the proof: 
\begin{enumerate}
\item First we compactify one of the directions ($x$) tangential to
the $p$-brane pair along a circle of radius $R$, and switch on
half a unit of Wilson line along $x$ on one of the $p$-branes.
This makes
the open string states with two ends lying on two different
branes anti-periodic under $x\to x+2\pi R$. As a result, the
zero $x$-momentum mode is absent from this open string sector, and at 
a critical
radius $R_c$, the lowest momentum mode of the tachyon associated with
such an open string becomes massless. 
The critical
radius turns out to be
half of the self-dual radius where the conformal field
theory associated with the scalar field $X$ develops an
$SU(2)_L\times SU(2)_R$ current algebra.

\item By exploiting the $SU(2)_L\times SU(2)_R$ current algebra, one
can show that the vertex operator associated with the massless
mode of the tachyon represents an exactly marginal deformation.
Switching on a vacuum expectation value $\alpha$ of this mode
gives rise to a solvable boundary conformal field theory (BCFT), 
so that the spectrum and correlation functions in this BCFT can be
calculated for all values of $\alpha$. In particular we show that
at $\alpha=1$, the BCFT is identical to the one describing
a D-$(p-1)$-brane
located on a circle of radius $R_c$ (with the directions
tangential to the D-$(p-1)$-brane being orthogonal to the
circle.) On the other hand, by
explicitly examining the form of the tachyon background, we find
that it corresponds to a tachyonic kink on a circle of radius
$R_c$. This allows us to identify the tachyonic kink on a circle
of radius $R_c$ to a D-$(p-1)$-brane on a circle of radius $R_c$.

\item In the final step, we increase the radius $R$ back to
infinity. Although at $R=R_c$ all values of $\alpha$ are
allowed, we find that as soon as $R$ increases beyond $R_c$, the
tachyonic mode which was massless at $R=R_c$ develops a tadpole
for a generic value of $\alpha$.
This shows that a generic $\alpha$ is no longer a solution of the
equations of motion. However, there are two inequivalent values
of $\alpha$ where the tadpole vanishes $-$ $\alpha=0$ and
$\alpha=1$. If we take the radius back to infinity keeping
$\alpha=0$, we recover the original D-$p$-brane pair, whereas
if we do so at $\alpha=1$, we expect to recover the
tachyonic kink solution. On the other hand, by examining the BCFT
corresponding to this configuration, we find that it describes a
D-$(p-1)$-brane of the bosonic string theory in 26 dimensional
Minkowski space. This proves the
identification of the tachyonic kink solution on the D-$p$-brane
pair with the D-$(p-1)$-brane.
\end{enumerate}

The paper is organised as follows. For simplicity of notation we
shall consider the case $p=1$, although generalisation to
arbitrary $p$ is completely straightforward. Also we shall work
in units where the fundamental string tension is given by
1/$(2\pi)$, {\it i.e.} $\alpha'=1$. In section \ref{s2} we
discuss some general properties of a pair of D-branes wrapped on
a circle, and carry out the step 1 of our analysis.
We also outline in slightly more detail the logic of
our analysis in steps 2 and 3. 
In sections \ref{s3} and \ref{s4} we explicitly carry out steps
2 and 3 of our analysis.   In section \ref{s5} we speculate on
the possibility of constructing the D0-brane as a tachyonic lump
on a single D-string. Again, this result can be easily
generalized to the case of a D-$p$-brane.
In section \ref{s6} we suggest a possible
reinterpretation of the results of sections \ref{s3} and \ref{s4}
which unifies the two apparently different descriptions of the
lower dimensional D-brane.

\sectiono{Pair of D-strings Wrapped on a Circle} \label{s2}

Our starting point will be a coincident pair of D-$p$-branes in
bosonic string theory. The open
strings living on this system are described by $2\times 2$ Chan
Paton (CP) factors. The massless sector of the open
string contains a U(2) gauge field living on the brane
world-volume. All open string states transform in the adjoint
representation of U(2) $-$ {\it i.e.} they are neutral under the
U(1) and transform in the singlet plus a triplet representation
of SU(2). States with CP factor $I$ (the $2\times
2$ identity matrix) are in the singlet representation of SU(2),
whereas those with CP factors $\sigma_i$ (the Pauli matrices) are
in the triplet representation of SU(2). 

The ground state of the open string in each sector describes a
tachyon field with
\be \label{e1}
m^2 = -1\, .
\ee
The first excited states correspond to the U(2) gauge fields as
well as $(25-p)$ massless scalars in the adjoint representation of
U(2). Higher excited states correspond to massive modes.
We shall denote the tachyonic state with CP factor $\sigma_1$ by
$T$. Let $V(T)$ denote the {\it classical} effective 
potential for this tachyon
obtained after integrating out the other massive string modes.
Since the SU(2) gauge transformation corresponding to the
group element $\exp(i\pi\sigma_3/2)$ takes $T$ to $-T$, $V(T)$
must be an even function of $T$.\footnote{If we also consider the
tachyonic modes coming from CP factors $\sigma_2$ and $\sigma_3$,
then there is an SU(2) triplet of tachyon field, and the
potential is invariant under the SU(2) transformation.} 
We shall argue shortly that $V(T)$
has (local) minimum at some values $\pm T_0$ such that
\be \label{e2}
V(\pm T_0) + 2 \TT_p = 0\, ,
\ee
where $\TT_p$ denotes the tension of the D-$p$-brane. Thus for
$T=\pm T_0$, the total energy density on the brane pair vanishes
and the system is indistinguishible from vacuum.

If $T_0\ne 0$, we can
define the tachyonic kink solution on the D-$p$-brane pair
as a solution of the equations of motion of classical open string
field theory, subject to the following conditions:
\begin{enumerate}
\item Only those open string fields which carry CP factors
$I$ and $\sigma_1$ are present as background.

\item The configuration is time independent, as well as
independent of $(p-1)$ of the spatial directions along the brane.

\item $T$ depends on the remaining spatial direction $-$ which
we shall denote by $x$ $-$ such that
\ben \label{e3}
T(x) &\rightarrow& T_0 \qquad \hbox{as} \qquad x\to\infty\, ,
\nonumber \\
&\rightarrow& -T_0 \qquad \hbox{as} \qquad x\to-\infty \, .
\een
\end{enumerate}
\begin{figure}[!ht]
\leavevmode
\begin{center}
\epsfbox{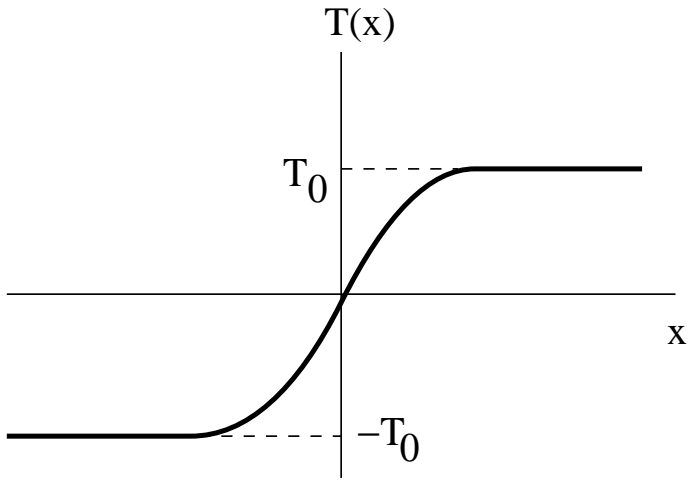}
\end{center}
\caption{The tachyonic kink on the pair of D-$p$-branes.}
\label{f1}
\end{figure}
This has been illustrated in Fig.\ref{f1} and describes a kink.
{}From eqs.\refb{e2} and \refb{e3} we see that the energy density
vanishes as $x\to\pm\infty$, and is concentrated near the core at
$x=0$. Thus this describes a $(p-1)$ dimensional brane. We shall
argue that this kink actually describes the D-$(p-1)$-brane of
the bosonic string theory.

Note that if \refb{e2} had not been true, the
tachyonic kink defined this way would have infinite tension when
regarded as a $(p-1)$-brane, since the energy density, integrated
along $x$, will not give a finite answer. Thus proving that the
tachyonic kink describes the D-$(p-1)$-brane $-$ which is known to
have finite tension $-$ automatically proves \refb{e2}.

A somewhat different scenario, which is also consistent with the
results of sections \ref{s3}, \ref{s4} will be suggested in
section \ref{s6}. In this description $T_0$ vanishes and the
D-$(p-1)$-brane is regarded as a tachyonic lump on a pair of
D-$p$-branes. The argument leading to the vanishing of $V(T_0)$
is still valid, but in this case the negative contribution to the
potential energy, cancelling the tension of the pair of
D-$p$-branes, comes from the vev of the tachyon associated with
the CP factor $I$.\footnote{Since $V(T)$ denotes the {\it
effective action} obtained after integrating out the other
fields, including the tachyon field from the identity sector, it
is automatically minimized with respect to this tachyon.}

\begin{figure}[!ht]
\leavevmode
\begin{center}
\epsfbox{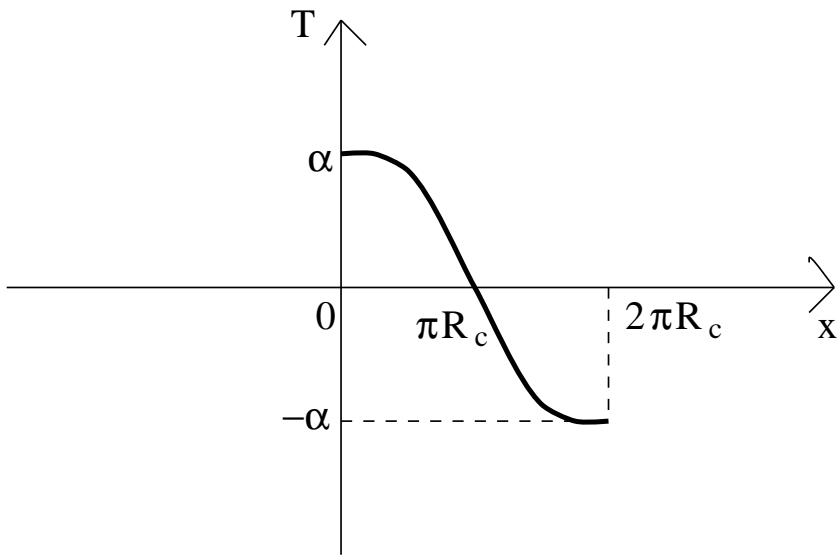}
\end{center}
\caption{The tachyonic kink on the pair of D-strings on a
circle.}
\label{f2}
\end{figure}
{}From now on we shall focus our attention
on the case $p=1$, and also compactify the
direction $x$ tangential to the D-string pair on a circle of
radius $R$. Associated with each D-string is a U(1) gauge field
which forms part of the full U(2) gauge group. We now switch on
half a unit of Wilson line on one of the D-strings. This breaks
the U(2) gauge symmetry to U(1)$\times$U(1). Presence of this
Wilson line does not affect the spectrum of open string states
with CP factors $I$ and $\sigma_3$, which represent strings with
both ends lying on the same D-string. But for open strings with
CP factors $\sigma_1$ and $\sigma_2$, corresponding to open
strings with two ends lying on two different D-strings, the wave
function is now required to be anti-periodic instead of periodic
under $x\to x+2\pi R$. In particular the tachyon $T$ coming from
CP factor $\sigma_1$ now has a mode expansion of the form:
\be \label{e4}
T(x) = \sum_{n\in Z} T_{n+{1\over 2}} e^{i(n+{1\over 2}){x\over
R}} \, .
\ee
The effective mass$^2$ of $T_{n+{1\over 2}}$ is given by:
\be \label{e5}
m_{n+{1\over 2}}^2 = {(n+{1\over 2})^2\over R^2}-1\, .
\ee
This shows that at the critical radius
\be \label{e6}
R_c={1\over 2}\, ,
\ee
the modes $T_{\pm{1\over 2}}$ become massless. In the next
section we shall show that the combination
\be \label{e7}
S\equiv T_{1\over 2} + T_{-{1\over 2}}
\ee
has vanishing potential as well, so that it represents an exactly
marginal deformation of the BCFT describing the D-string pair. We
can parametrise a general vacuum expectation value (vev) of this
marginal operator as:\footnote{Once the interactions are
taken into account, the higher modes of the tachyon will also
acquire non-zero vev, but $T(x)$ will continue to have the shape
of a kink.}
\be \label{e8}
T_{1\over 2}+T_{-{1\over 2}}=\alpha, \qquad 
T_{1\over 2}-T_{-{1\over 2}}=0, \qquad 
T_r=0 \quad \hbox{for} \quad |r|>{1\over 2}\, .
\ee
{}From \refb{e4} we see that this corresponds to
\be \label{e9}
T(x) = \alpha \cos{x\over 2R_c}=\alpha\cos x\, .
\ee
This has been plotted in Fig.\ref{f2}, and clearly represents a
tachyonic kink on a circle of radius $R_c={1\over
2}$.\footnote{It actually represents an anti-kink, but since
bosonic D-branes do not carry any charge, there is no distinction
between a kink and an anti-kink.}

After switching on the tachyon vev, we would like to take the
radius back to infinity. We shall see in section \ref{s4} that as
soon as we make $R>R_c$, the field $S$ develops a tadpole
for a generic value of $\alpha$.
This is not surprising since $S$ represents an exactly marginal
deformation only for $R=R_c$. However, we find that the tadpole
vanishes for two inequivalent values of $\alpha$, namely, $\alpha=0$
and $\alpha=1$. Thus it is natural to identify the $\alpha=1$
configuration as the tachyonic kink on a circle of radius $R$. We
show in section \ref{s4} that the BCFT corresponding to this
configuration is identical to the one describing a D-particle on
a circle of radius $R$. This equivalence continues to hold in the
$R\to\infty$ limit as well.

\begin{figure}[!ht]
\leavevmode
\begin{center}
\epsfbox{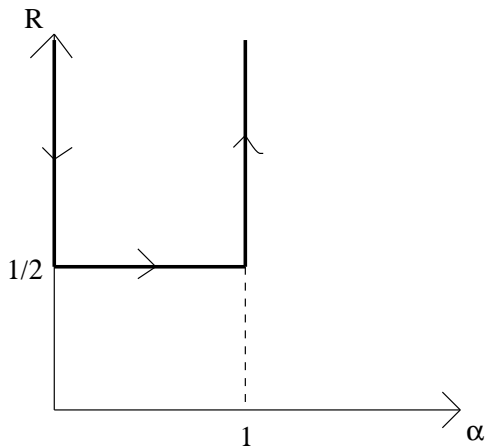}
\end{center}
\caption{The marginal flow in the $R-\alpha$ plane interpolating
between a pair of D-strings and a D-particle.}
\label{f3}
\end{figure}
Thus we see that there is a series of marginal deformations which
relate a pair of D-strings to the D-particle. In the $R-\alpha$
plane, this marginal flow has been depicted in Fig.\ref{f3}.

As a consistency check for this proposal, let us note that the
mass of the wrapped pair of D-strings on a circle of radius
${1\over 2}$ is given by,
\be \label{e34}
2 \cdot 2\pi \cdot {1\over 2} \cdot \TT_1\,  .
\ee
Using the relation\cite{NOTES}
\be \label{e35}
\TT_p={1\over 2\pi} \TT_{p-1}\, ,
\ee
wee see that \refb{e34} is identical to the mass $\TT_0$ of a
single D-particle.

\sectiono{CFT at $R={1\over 2}$} \label{s3}

The world-sheet field that will be important for our analysis is
the scalar field $X$ associated with the coordinate along the
D-string. Besides this, there are 25 other scalar fields
$X^0,\ldots X^{24}$, and the ghost fields $b,c,\wb b,\wb c$.
Although $X$ has radius ${1\over 2}$, it will be more convenient
for our analysis to regard this as a scalar field of radius 1,
and then mod out the theory by the transformation:
\be \label{e11}
h_X: X\to X+\pi\, 
\ee
All closed string states, as well as open string states with CP
factors $I$ and $\sigma_3$ are required to be even under $h_X$,
whereas open string states with CP factors $\sigma_1$ and
$\sigma_2$ are required to be odd under $h_X$. 

We shall express $X$
as a sum of the left and the right-moving parts:
\be \label{e12}
X\equiv X_L+X_R\, ,
\ee
with $h_X$ acting on $X_L$ and $X_R$ as
\be \label{e12a}
h_X:\quad X_L\to X_L+{\pi\over 2}, \quad X_R\to 
X_R+{\pi\over 2}\, .
\ee
At unit radius, the CFT describing $X$ possesses an
SU(2)$_L\times$SU(2)$_R$ current algebra. This allows us to introduce
two other bosons:
\be \label{e13}
\phi\equiv \phi_L+\phi_R, \qquad \phi'\equiv \phi'_L+\phi'_R\, ,
\ee
through the relations:
\be \label{e14a}
e^{2iX_L}=\p\phi_L + i\p\phi'_L, \qquad
e^{2iX_R}=\p\phi_R + i\p\phi'_R\, .
\ee
The set of left-moving currents $(\p X_L,\p\phi_L,\p\phi'_L)$
transform as a triplet of SU(2)$_L$; similarly the right-moving
currents transform as a triplet of SU(2)$_R$. {}From \refb{e14a}
we can also write down the SU(2) rotated version of these
relations:\footnote{There is some freedom in writing down these
relations, since eq.\refb{e14a} only defines $\phi$,
$\phi'$ up to a constant shift.}
\ben \label{e14b}
&&e^{2i\phi_L}=\p X_L - i\p \phi'_L, \qquad
e^{2i\phi_R}=\p X_R - i\p \phi'_R, \nonumber \\
&&e^{2i\phi'_L}=\p X_L + i\p \phi_L, \qquad
e^{2i\phi'_R}=\p X_R + i\p \phi_R\, .
\een

We have already introduced the transformation $h_X$. We now
introduce a new transformation:
\be \label{e16}
g_X: \quad X \to -X\, .
\ee
We also introduce the transformations $g_\phi$, $h_\phi$,
$g_{\phi'}$ and $h_{\phi'}$ in an identical manner. Then, using
eqs.\refb{e14a}, \refb{e14b} we see that,
\be \label{e17}
h_X=g_{\phi}=g_{\phi'}, \qquad g_X=h_\phi =h_{\phi'}g_{\phi'}\, .
\ee

We shall use the convention where the open string world-sheet
corresponds to the upper half plane spanned by a complex
coordinate $z$, and the right-moving currents are holomorphic in
$z$. The operator product expansion of the currents are given by,
\be \label{e18}
\p X_R(z) \p X_R(w) \simeq \p \phi_R(z)\p \phi_R(w) \simeq
\p\phi'_R(z) \p\phi'_R(w) \simeq -{1\over 2(z-w)^2}\, ,
\ee
where $\simeq$ denotes equality up to non-singular terms. There
are similar relations involving the left-moving currents.

Since both the D-strings are stretched along $x$, we impose
Neumann boundary condition on $X$:
\be \label{e19}
(X_L)_B=(X_R)_B\equiv {1\over 2} X_B\, ,
\ee
where the subscript $B$ denotes boundary value. Using
eq.\refb{e14a} this translates into Neumann boundary condition on
$\phi$ and $\phi'$:
\be \label{e20}
(\phi_L)_B=(\phi_R)_B\equiv {1\over 2}\phi_B, \qquad 
(\phi'_L)_B=(\phi'_R)_B\equiv {1\over 2}\phi'_B\, .
\ee
The vertex operator for the mode $T_{n+{1\over 2}}$ defined in
\refb{e4} is given by:
\be \label{e21}
V_{n+{1\over 2}} = i e^{2i(n+{1\over 2}) X_B}\otimes \sigma_1\, .
\ee
Thus the vertex operator for $S\equiv(T_{1\over 2}+T_{-{1\over
2}})$ is given by:
\be \label{e22}
V_S=i(e^{iX_B}+e^{-iX_B})\otimes\sigma_1 =
i\p\phi_B\otimes\sigma_1\, .
\ee
In deriving the right hand side of \refb{e22} we have used
eq.\refb{e14a}. At this stage,
the overall normalization of $V_S$ is arbitrary. From \refb{e22}
we see that $V_S$ represents the vertex operator of a zero
momentum gauge field $\AAA_\phi$ along $\phi$ with CP factor
$\sigma_1$, $-$ in other words a Wilson line along $\phi$. This
clearly is an exactly marginal deformation.
We parametrize it by a parameter $\alpha$, normalized such that
it corresponds to inserting an operator
\be \label{e23}
\exp (i{\alpha\over 4}\sigma_1 \ointop \p_t \phi_B dt)\, ,
\ee
on the boundary of the world-sheet. $t$ denotes a parameter along
this boundary.

We shall now study the effect of switching on this Wilson line on
the open string states. Using the relations
\be \label{e24}
[\sigma_1,I]=0=[\sigma_1,\sigma_1]\, ,
\ee
we see that open string states with CP factors $I$ and $\sigma_1$
are neutral under this gauge field, and hence the spectrum of
these open string states is not modified upon switching on the
tachyon vev. On the other hand, since
\be \label{e25}
[\sigma_1,\sigma_3\mp i\sigma_2]=\pm 2 
(\sigma_3\mp i\sigma_2)\, ,
\ee
open string states in these sectors carry charge $\pm 2$ under
$\AAA_\phi$. Thus the momentum $p_\phi$ along the $\phi$
direction gets shifted;
\be \label{e26}
p_\phi\to p_\phi \pm {\alpha\over 2}\, .
\ee
Eq.\refb{e26} can be restated by saying that the $h_\phi$ quantum
number of the state is multiplied by a factor:
\be \label{e28}
\exp(\pm i\pi{\alpha\over 2})\, .
\ee
{}Using the identification of $h_\phi$ with $g_X$ (eq.\refb{e17})
and the fact that the open string spectrum has no $g_X$
projection, we see that before switching on the tachyon vev, the
open string states in each CP factor contains both $h_\phi$ even
and $h_\phi$ odd states. If $\alpha=2$, then, as seen from
eq.\refb{e28}, the $h_\phi$ quantum numbers get multiplied by
$-1$. But this means that the complete spectrum of open strings
in each CP sector remains unchanged. Thus the BCFT at $\alpha=2$
is equivalent to that at $\alpha=0$, and we conclude that
$\alpha$ is a periodic variable with period 2. 

Although at the critical radius $R=R_c$, all values of $\alpha$
describe consistent BCFT, in anticipation of the results of the
next section let us pay special attention to the spectrum at
$\alpha=1$. We begin by tabulating the $h_\phi$ and $g_\phi$
eigenvalues of various open string states at $\alpha=0$:

\begin{center}
\begin{tabular}{|c|c|c|}
\hline
CP factors & $g_X=h_\phi$ & $h_X=g_\phi$ \\
\hline
$I$ & $\pm 1$ & 1 \\
\hline
$\sigma_1$ & $\pm 1$ & $-1$ \\
\hline
$\sigma_2$ & $\pm 1$ & $-1$ \\
\hline
$\sigma_3$ & $\pm 1$ & 1 \\
\hline
\end{tabular}
\medskip

Table 1: Spectrum at $\alpha=0$
\end{center}

For $\alpha=1$, the quantum numbers in sectors $I$ and $\sigma_1$
remain unchanged, but the $h_\phi$ eigenvalues in sectors
$\sigma_3\mp i\sigma_2$ get multiplied by $\pm i$. Thus the
above table is modified to:

\begin{center}
\begin{tabular}{|c|c|c|}
\hline
CP factors & $h_\phi=h_{\phi'}g_{\phi'}$ & $g_\phi=g_{\phi'}$ \\
\hline
$I$ & $\pm 1$ & 1 \\
\hline
$\sigma_1$ & $\pm$ 1 & $-1$ \\
\hline
$\sigma_2$ & $\pm i$ & $-1$ \\
\hline
$\sigma_3$ & $\pm i$ & 1 \\
\hline
\end{tabular}
\medskip

Table 2: Spectrum at $\alpha=1$
\end{center}

{\it Combining the open string spectrum from all sectors} 
we see that
{\it in the Fock space},
\begin{itemize}
\item there is no $g_{\phi'}$ projection, and
\item all $\phi'$ momentum of the form:
\be \label{e29}
p_{\phi'} = {n\over 2}, \qquad n\in Z\, ,
\ee
are allowed.
\end{itemize}
Let us define a new field $\phi''$ related to $\phi'$ by
T-duality transformation:
\be \label{e30}
\phi''_L=\phi'_L, \qquad \phi''_R=-\phi'_R\, ,
\ee
If $w_{\phi''}$ denotes the winding charge along $\phi''$
(defined as $\Delta\phi''/2\pi$), and
$g_{\phi''}$ denotes the transformation $\phi''\to -\phi''$, then
we have the relations:
\be \label{e31}
p_{\phi'}=w_{\phi''}, \qquad g_{\phi'}=g_{\phi''}\, .
\ee
We now note that in $\phi''$ coordinate,
\begin{enumerate}
\item The boundary condition \refb{e20} takes the form:
\be \label{e32}
(\phi''_L)_B=-(\phi''_R)_B\, .
\ee
In other words, we have Dirichlet boundary condition along
$\phi''$.

\item The full spectrum of open strings from all CP sectors has
no $g_{\phi''}$ projection in the Fock space.

\item The $\phi''$ winding charge is quantized as:
\be \label{e33}
w_{\phi''} = {n\over 2}\, , \qquad n\in Z\, .
\ee
\end{enumerate}
This is precisely the spectrum of open strings living on a single
D-particle on a circle of radius ${1\over 2}$.

One can ask whether the 
interaction among these open strings is also identical to that
among open
strings living on a D-particle. The difference between the
interaction rules in the two theories comes from the CP factors.
In the BCFT under consideration various open string vertex
operators are accompanied by CP factors, and there could be extra
selection rules, as only those products of CP factors with
non-vanishing trace will give non-zero amplitude. This requires,
for example, that either each $\sigma_i$ come in pairs, or they
appear in the combination $\sigma_1\sigma_2\sigma_3$. But upon
examining the $h_{\phi'}$ and $g_{\phi'}$ quantum numbers carried by
these states, we see that these selection rules are automatically
imposed by $h_{\phi'}$ and $g_{\phi'}$ conservation laws, which are
present also for the open string living on D-particle on a
circle. This shows that non only the spectrum, but also the
correlation functions of the BCFT obtained here agrees with those
in the BCFT describing
D-particle on a circle of radius ${1\over 2}$.

This shows the equivalence between the tachyonic kink on a pair
of D-strings on a circle of radius ${1\over 2}$, and the
D-particle on a circle of the same radius.
In the next section we shall show that this identification of the
two boundary conformal field theories persists even when we
increase the radius back to $\infty$.

\sectiono{Taking the Radius Back to $\infty$} \label{s4}

The analysis in this section will follow closely that of
ref.\cite{SPINOR}. The effect of increasing the radius is
achieved by perturbing by the closed string vertex operator:
\be \label{e36}
\int d^2 z \p X_L \p X_R \equiv \int d^2 z V_r\, .
\ee
First we shall show that even at first order in $(R-R_c)$, the
field $S$ develops a tadpole at a generic value of $\alpha$. This
one point function is proportional to the two point function at
$R=R_c$ of
the open string vertex operator $V_S$ and the closed string
vertex operator $V_r$ in the presence of the tachyonic
background parametrised by $\alpha$. This is
given by\footnote{Ghost number
conservation requires that we also insert appropriate number of
ghost fields in the vertex operators. But the correlation
function factors into a matter part and the ghost part, and
non-trivial information comes from analysing the matter part.
Hence we focus on the matter part of the correlator.}
\be \label{e37}
\langle V_S V_r\rangle_\alpha 
\propto Tr_{CP}\Big[\Big\langle \p X_L\p X_R(P) \p 
\phi_B(Q)\otimes\sigma_1
\exp(i{\alpha\over 4}\sigma_1\ointop \p_t\phi_B dt) 
\Big\rangle\Big]\, .
\ee
In the left hand side of this equation $\langle~\rangle_\alpha$
denotes the correlation function in the presence of the tachyonic
background. In the right hand side $\langle~\rangle$ denotes the
correlation function at $\alpha=0$; the effect of the tachyonic
background has been taken into account by explicitly putting the
exponential factor inside the correlator. $P$ denotes a point in
the interior of the world-sheet, $Q$ denotes a point on the
boundary, and $Tr_{CP}$ denotes trace over the Chan Paton
factors. By explicitly carrying out the trace over the CP factors,
and using the relations \refb{e14b}, we can rewrite the right
hand side of \refb{e37} as
\be \label{e39}
{1\over 2} \Big\langle \sin\big({\alpha\over 2}\pi w_\phi\big) \Big(
e^{2i(\phi_L+\phi_R)}+
e^{-2i(\phi_L+\phi_R)}+
e^{2i(\phi_L-\phi_R)}+
e^{-2i(\phi_L-\phi_R)}\Big)(P)\p\phi_B(Q)\Big\rangle \, ,
\ee
where,
\be \label{e38}
w_\phi={1\over 2\pi} \ointop \p_t\phi_B dt\, .
\ee
$w_\phi$ measures the total $\phi$ winding charge carried by the
closed string vertex operators inserted in the interior. Since
$\exp(\pm 2i(\phi_L+\phi_R))$ has $w_\phi=0$, whereas
$\exp(\pm 2i(\phi_L-\phi_R))$ has $w_\phi=\pm 2$, we can rewrite
\refb{e39} as
\be \label{e40}
{1\over 2}(\sin{\alpha\pi})
\Big\langle\Big(e^{2i(\phi_L-\phi_R)}-
e^{-2i(\phi_L-\phi_R)}\Big)(P)\p\phi_B(Q)\Big\rangle \, .
\ee
Using the fact that $\phi$ satisfies Neumann boundary condition,
one can easily show that the correlation function appearing in
\refb{e40} is non-zero. This shows that the one point function of
$S$ for $R>R_c$ is non-zero. It vanishes at $\alpha=0$ and at
$\alpha=1$, as stated earlier.

We shall now focus our attention on the $\alpha=1$ point, and
analyse a general correlation function of open string vertex
operators at a general radius $R$. This would require summing
over
arbitrary number of insertions of $V_r$ with appropriate weight
factors. Thus a typical correlation function to be analysed has
the form:
\be \label{enew1}
\Big\langle \prod_i V^i_{open}(Q_i) \prod_m V_r(P_m)
\exp\Big(i{\pi\over 2} w_\phi \sigma_1\Big)\Big\rangle \, ,
\ee
where $w_\phi$ has been defined in eq.\refb{e38}. Here $Q_i$ are
points on the boundary of the world-sheet and $P_m$ are points in
the interior. The effect of the exponential factor on the open
string vertex operators is to shift their $\phi$ momentum, which
has already been taken into account in the previous section. Thus
we can interprete $w_\phi$ appearing in \refb{enew1} as the sum
of the $\phi$ winding charges of all the closed string vertex
operators inserted in the interior. Expressing $V_r$ as
\be \label{e43}
V_r 
= {1\over 4} \Big( e^{2i(\phi_L+\phi_R)}+ e^{-2i(\phi_L+\phi_R)}+
e^{2i(\phi_L-\phi_R)}+
e^{-2i(\phi_L-\phi_R)}\Big)\, ,
\ee
we see that the first two terms carry $w_\phi=0$ whereas the
third and the fourth terms carry $w_\phi=\pm 2$. Since all terms
have $w_\phi$ even, we have the relation:
\be \label{e44}
\exp(i{\pi\over 2}w_\phi\sigma_1) = (-1)^{w_\phi/2}\, .
\ee
This transforms $V_r$ given in \refb{e43} to
\be \label{e45}
{1\over 4} \Big( e^{2i(\phi_L+\phi_R)}+ e^{-2i(\phi_L+\phi_R)}-
e^{2i(\phi_L-\phi_R)}- e^{-2i(\phi_L-\phi_R)}\Big)
=-\p\phi'_L\p\phi'_R=\p\phi''_L\p\phi''_R\, ,
\ee
where $\phi''$ has been defined in eq.\refb{e30}.
Perturbing by the operator $\p\phi''_L\p\phi''_R$ has the effect
of increasing the $\phi''$ radius (or, equivalently, decreasing
the $\phi'$ radius). Thus we see that the effect of increasing
$R$ in the presence of a tachyon background is achieved by
increasing the $\phi''$ radius in the same proportion, and
ignoring the tachyon background. Since for $R=R_c={1\over 2}$,
the open string spectrum corresponds to a D-particle on the
$\phi''$ circle of radius ${1\over 2}$, we see that if we
increase $R$ to $\lambda\cdot {1\over 2}$, this would correspond
to a D-particle on the $\phi''$ circle of radius $\lambda\cdot
{1\over 2} =R$. As $R\to\infty$, this corresponds to a D-particle
in the (25+1) dimensional Minkowski space.

This proves the equivalence of the BCFT describing the tachyonic
kink solution on a pair of D-strings,
and that describing a D-particle. Note that the marginal
deformation interpolating between the two configurations does
not involve the fields $X^0,\ldots X^{24}$ at any step. Thus by
putting Neumann boundary condition on $(p-1)$ of the fields
$X^1,\ldots X^{24}$ we can make our initial configuration into a
pair of D-$p$-branes, and the final configuration into a
D-$(p-1)$-brane.

\sectiono{D-particle as a Tachyonic Lump on a Single D-string?}
\label{s5}
\begin{figure}[!ht]
\leavevmode
\begin{center}
\epsfbox{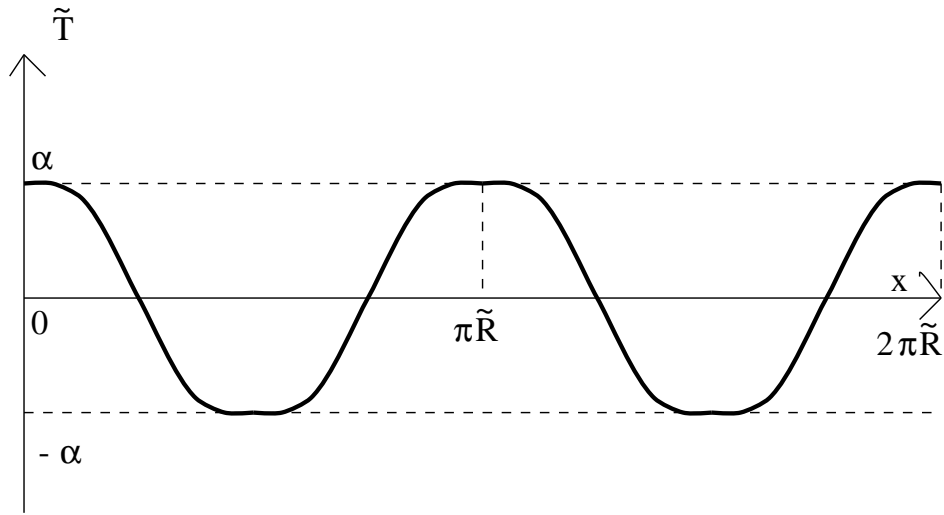}
\end{center}
\caption{The tachyon field on the D-string which produces a pair
of D-particles.}
\label{f4}
\end{figure}

Following ref.\cite{NOSU}, we can study a T-dual
version of our analysis.\footnote{Many of the results of this section
have been discussed earlier in refs.\cite{RECK,CAL,POL}.}
In this case the massless mode of the
tachyon interpolates between a pair of D-particles situated at
diametrically opposite points on a circle of radius $\wt R=2$,
and a single D-string wrapped on a circle of the same radius.
Running the marginal flow backwards, we can conclude that tachyon
condensation on a D-string on a circle of radius $\wt R$ produces
a pair of D-particles at diametrically opposite points on the
same circle. From the point of view of the D-string, the
particular mode which condenses corresponds to:
\be \label{ef1}
\wt T(x) = \wt\alpha\cos x = \wt \alpha\cos{2x\over \wt R}\, ,
\ee
where $\wt T$ is the tachyonic field living on the D-string. This
has been plotted in Fig.\ref{f4}. This can be viewed as a pair of
lumps on the circle, one spanning the range $0\le x\le \pi\wt R$,
and the other spanning the range $\pi\wt R\le x\le 2\pi\wt R$.
Although this configuration is symmetric under $\wt T\to -\wt T$
(together with a translation along $x$), this symmetry will be
destroyed once interactions are taken into account.

\begin{figure}[!ht]
\leavevmode
\begin{center}
\epsfbox{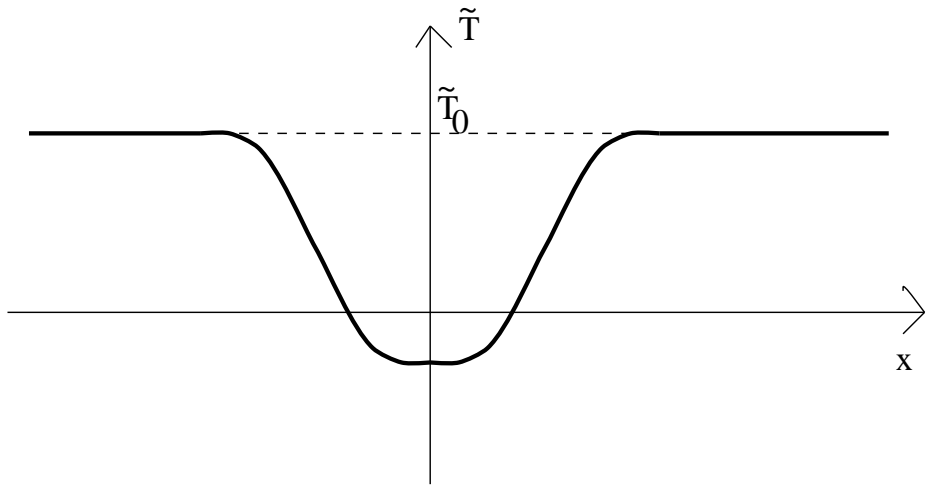}
\end{center}
\caption{The tachyonic lump on a D-string, representing a single
D-particle.}
\label{f5}
\end{figure}

This suggests that we can identify a single D-particle in 26
dimensional Minkowski space as a
tachyonic lump on an infinite D-string as shown in Fig.\ref{f5}. 
In this
figure $\wt T_0$ denotes the minimum of the tachyonic potential
$\wt V(\wt T)$ satisfying,
\be \label{ef2}
\wt V(\wt T_0) + \TT_1 = 0\, .
\ee
This relation guarantees that the lump has a finite mass.

This analysis can be generalised by starting from a D-string
compactified on a circle of radius $n$ for any integer $n$
(including 1), and condensing the mode proportional to $\cos x$.
This will produce $n$ lumps on the circle, which can be
identified as $n$ uniformly spaced D-particles on a circle of
radius $n$\cite{RECK,CAL,POL}.

\sectiono{Reconciling the Two Different Descriptions of D-branes}
\label{s6}

\begin{figure}[!ht]
\leavevmode
\begin{center}
\epsfbox{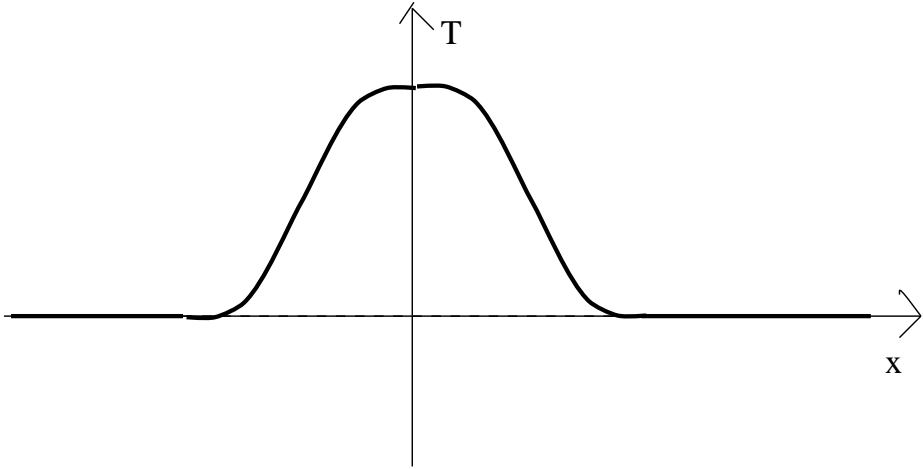}
\end{center}
\caption{Coefficient of $\sigma_1$ in eq.\refb{ex4}.}
\label{f6}
\end{figure}

The analysis of the previous sections suggests two different ways
of viewing a D0-brane (D-$(p-1)$-brane), $-$ as a kink on a
D-string anti-D-string (D-$p$-brane anti-D-$p$-brane) pair, and
also as a lump on a single D-string (D-$p$-brane). In this
section we shall show that we can reinterprete the results of
sections \ref{s3} and \ref{s4} in such a way that the second
description can explain all the results of this paper. To do this
let us denote by $f_{L}(x)$ the function shown in Fig.\ref{f5}.
Thus on a single D1-brane, the configuration
\be \label{ex1}
\wt T(x)=f_L(x)
\ee
denotes a D0-brane. On the other hand, $\wt T =\wt T_0$ denotes
the vacuum solution. Thus if $\wh T$ denotes the $2\times 2$
matrix valued tachyon field on a pair of coincident D-branes,
then the configuration:
\be \label{ex2}
\wh T(x) = \pmatrix{\wt T_0 & \cr & f_L(x)}
\ee
will denote a single D0-brane. \refb{ex2} can be rewritten
as,
\be \label{ex3}
\wh T(x) = {1\over 2} (f_L(x)+\wt T_0) I +{1\over 2} (\wt T_0 -f_L(x)
) \sigma_3\, .
\ee
Finally, using the U(2) gauge symmetry on the coincident D-string
pair, we can replace $\sigma_3$ in eq.\refb{ex2} by $\sigma_1$.
This gives,
\be \label{ex4}
\wh T(x) = {1\over 2} (f_L(x)+\wt T_0) I +{1\over 2} (\wt T_0 -
f_L(x)) \sigma_1\, .
\ee
The coefficient of $\sigma_1$ is the tachyon field $T(x)$ of
sections \ref{s2}-\ref{s4}. From Fig.\ref{f5} we see that this
has the form given in Fig.\ref{f6}. This appears to be a lump
rather than a kink, as suggested by the analysis of sections
\ref{s4} and \ref{s5}.
\begin{figure}[!ht]
\leavevmode
\begin{center}
\epsfbox{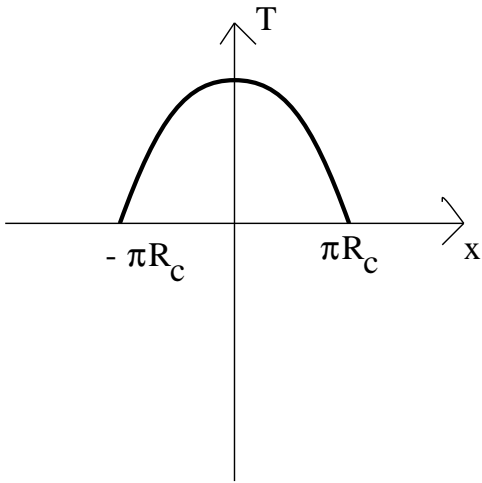}
\end{center}
\caption{A redrawing of the configuration of Fig.\ref{f2}.}
\label{f7}
\end{figure}
\begin{figure}[!ht]
\leavevmode
\begin{center}
\epsfbox{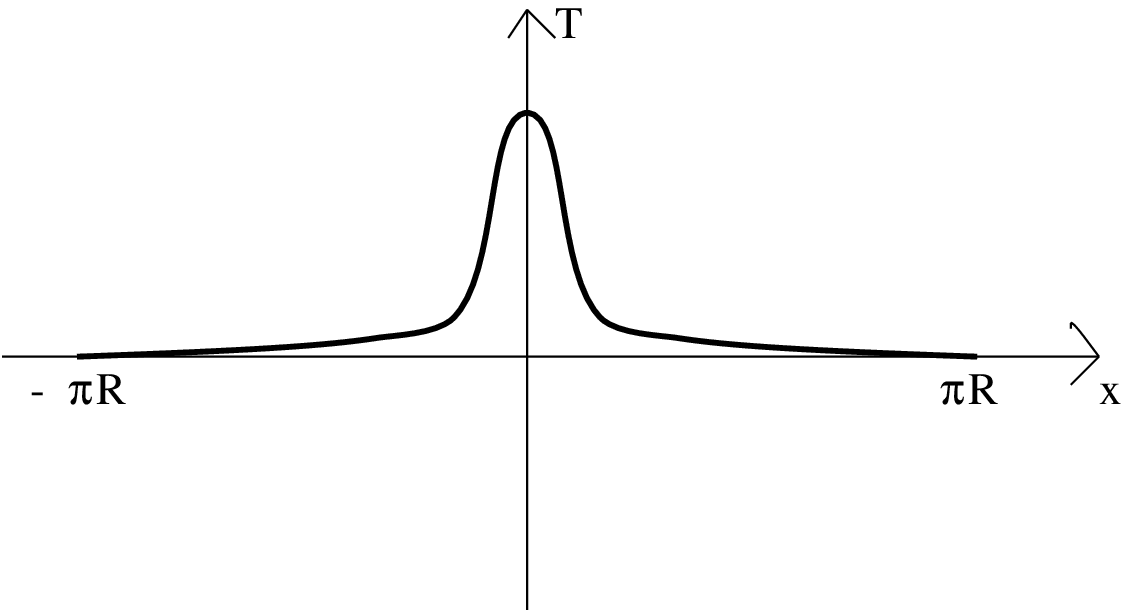}
\end{center}
\caption{A possible $R\to\infty$ limit of the configuration of
Fig.\ref{f7}. This provides an alternative to the possibility
shown in Fig.\ref{f1}.}
\label{f8}
\end{figure}

We shall now show that it is in principle possible to
reinterprete the tachyonic configuration of sections \ref{s4},
\ref{s5} to be a lump rather than a kink. For this we need to
note that Fig.\ref{f2} can be
redrawn as Fig.\ref{f7} by changing the range of $x$
parametrizing $S^1$ from $(0,2\pi R_c)$ to $(-\pi R_c,\pi R_c)$.
In this form the tachyon configuration does appear as a lump
rather than a kink.
The main question is: as we
take the $R\to\infty$ limit, does most of the circle get covered
by the configuration $T\simeq\pm T_0$ for some $T_0\ne 0$
(shown in Fig.\ref{f1}), 
or does it get covered by the
configuration $T\simeq 0$ (as shown in Fig.\ref{f8})? 
In the first case the configuration is to be
interpreted as a kink, whereas in the second case
it is to be interpreted as a lump as shown in Fig.\ref{f6}.

Note that in the second case
$T(x)\to 0$ as $x\to \pm\infty$. In the case of the D-string -
anti-D-string pair of type IIB string theory this was unacceptable,
since for $T=0$ the energy per unit length of the pair
is finite, and hence the resulting solution would not represent a
localised lump of energy. In the present case however there is a
tachyon from the identity sector as well, and
condensation of this tachyon can make the energy
density vanish far away from the core even though the tachyon
associated with the CP factor $\sigma_1$ has no vev in this
region. Indeed, this is
precisely what happens for the configuration of eq.\refb{ex4}.
Thus if we reinterprete
the results of section \ref{s3} and \ref{s4} this way, all the
results of this paper
would be consistent with the hypothesis that the tachyonic lump on a
single D-string represents a D0-brane.

\end{document}